# A new method to calculate the Total Fertility Rate from the number of birth

*Weidong Huang**

*Anhui Key Laboratory of Polar Environment and Global Change, School of earth and space science, University of Science and Technology of China*

*96 Jinzhai Road, Hefei, Anhui 230026*

**Corresponding author: Huang, email: huangwd@ustc.edu.cn (CA)*

**Abstracts**

**The standard methods to calculate the Total Fertility Rate require the reliable age-specific fertility rate including birth data and the related age-specific women's population data. Historically, the number of births was often not counted according to the age of the mother, so it is difficult to estimate the historical total fertility rate with the standard methods. Many empirical methods have been proposed, but their application is limited to specific period and place. This paper deduces a new method for calculating the total fertility rate from the number of birth and the population of women at childbearing age, can be applied to most of cases. The relative error is usually less than 5%. It is easier to calculate TFR, and may be applied to obtain more TFRs for the history.**

## 1 Introduction

The total fertility rate (TFR) is a comprehensive index of women's fertility levels and is the sum of the fertility rates of women at various childbearing age in a place (usually 15 to 49 years old). If the age-specific fertility rate for the year is maintained forever, the average fertility rate of women will be equal to the total fertility rate for the year. Therefore, the total fertility rate can be regarded as the average fertility rate of women in the year, which is a very important demographic parameter.

According to the definition of TFR, the calculation requires a large amount of high-quality data, especially the reliable age-specific fertility rate including birth data and the related age-specific women's population data. Historically, the number of births was often not counted according to the age of the mother (Preston, Heuveline and Guillot 2001), so it is difficult to estimate the total fertility rate according to the standard method Almost all demographers and officials acknowledge that there is a serious underreporting of the birth population registration in China, resulting in a low number of births, resulting in a low fertility rate. For example, studies have shown that the 0-1 year old population reported has a false of more than 20% negative rate in the 2000 census (Wang and Ge 2013). To develop a method based on a small amount of high-quality data to calculate the total fertility rate has become a hot topic in demographic research of China (Zhang 2006).

The empirical relationship between birth rate and general fertility rate had been discussed (Entwisle 1981). A linear empirical relationship between total fertility rate and crude birth rate is first established (You 1990) based on the data from the third population census in Guangxi Province of China. It has been proposed (Smith 1992) that the total fertility rate and the general fertility rate are directly proportional to the crude birth rate. Based on China's census data, a complex relationship is constructed (Wang 2002) between birth population and age-specific fertility rate and number of women, and the model parameters is determined through genetic algorithm, trying to improve the accuracy of fertility calculation. In 2006, the empirical relationship between total fertility rate and the general fertility rate has also been established (Zhang 2006). Based on the R method(Preston and Coale 1982), a method for calculating the total fertility rate based on net population reproductivity is also proposed (Preston et al. 2001). In 2008, it has been applied to calculate the fertility level in China from 1990 to 2000, including TFR(Cai 2008); the fertility level and total fertility rate in China from 2000 to 2010(zhao 2015). In 2018, the census data and "small census" data are applied to estimate the ratio of birth rate to total fertility rate, which remained constant in the short term (QIAO and ZHU 2018).

However, all these methods need to determine some key empirical parameters in the model based on the existing data. Generally, they can only be applied to a certain time period in a certain place, and cannot be extended to other places or at other times, so their application are limited, not able to applied for historical data.

In order to break through this limitation, this paper deduces a general method for estimating the TFR based on the total number of births and the total number of women of childbearing age. Using the fertility data of China and the United States, the calculation error of the method was evaluated, and the source of the error was theoretically analyzed. The method error and the scope of application were analyzed and estimated.

**2 method for deducing equation to calculate the total fertility rate**

According to the definition of **the total fertility rate** (TFR) in demography (Preston et al. 2001), it can be calculated as:

$$TFR=\Sigma f_x \quad (1)$$

Where fx is the age-specific fertility rate of x-year-old women of childbearing age, which can be calculated as:

$$fx=B_x/W_x \quad (2)$$

Where $B_x$ is the number of children born by x-year-old women, $W_x$ is the number of women aged x.

Because it is more difficult to obtain age-specific fertility data at each age for women of childbearing age, statistical age-specific fertility data is usually grouped by every 5 years old in the United States. In that case:

$$TFR=n*\Sigma\, {}_nf_x \quad (3)$$

Where n is 5. This method is abbreviated as the five-age method in this paper. This method reduces the amount of data required compared to the standard methods. In general, the women of childbearing age range in age from 15 to 49 years, for a total of 35 years. Let n=35, then the above equation becomes:

$$TFR=35*{}_{35}f_x=35*B/W \quad (4)$$

Where B is the total number of people born in the year, and W is the total number of women at childbearing age in the year. According to the above equation, calculating the total fertility rate requires only the number of births and the number of women at childbearing age.

Because this method calculates TFR based on the total birth population, we refer to it as the total birth number method. Since women at childbearing age have a small number of births in the youngest and oldest years, the births of these ages can be ignored, and the parameter n can be reduced (the number of groups to be considered is reduced) if the childbearing age range is changed.

In a short period of time, the proportion of W in the total population usually varies a little,  if K is defined as the proportion of women at childbearing age in the total population:

$$K=W/P \tag{5}$$

Where P is the total population. K can be approximately as constant in a certain period of time. If equation (5) is substituted into equation (4),

$$TFR=35*B/(KP)=35/K*CBR \tag{6}$$

Where CBR is the crude birth rate. That means the new TFR calculation method deduced in this paper is consistent with the former study (QIAO and ZHU 2018) which proposed that a ratio of birth rate to total fertility rate remained constant in the short term, and shows that the TFR is proportional to the crude birth rate, the meaning of the proportional coefficient is also given, which proposes a good method to estimate the coefficient, so the equation (6) is changed from the empirical method to a general method.

**3 Validation and error assessment of the Method**

Using the data of the age-specific women and the corresponding fertility population, the total fertility rate can be calculated with the standard method or the present new method. Then the error of the present method can be estimated by compare the results of these two methods.

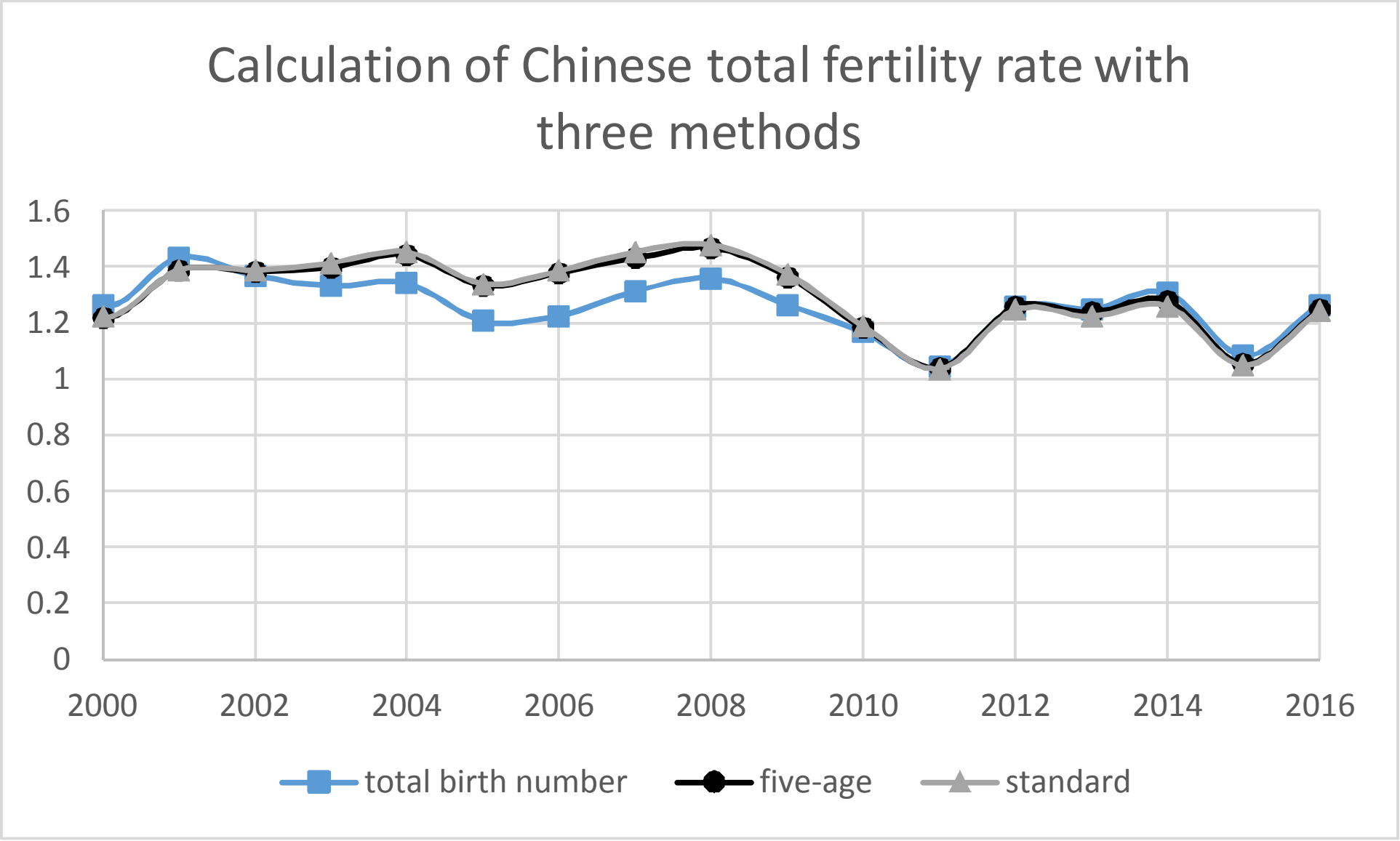

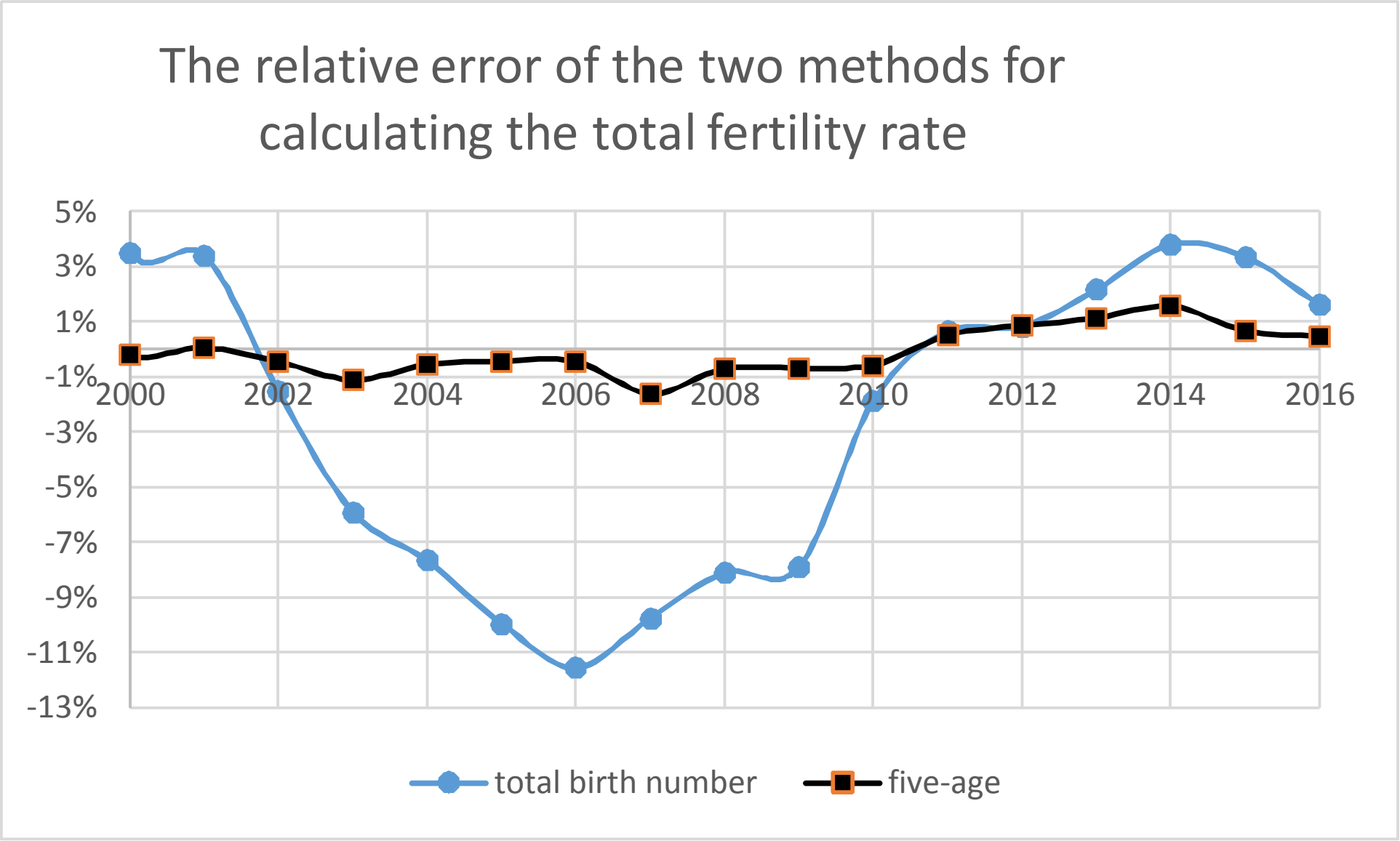

Figure 1, Comparison of standard method, five-age method and total birth number method to calculate the total fertility rate in China from 2000 to 2016

Upper: the total fertility rate calculated by the three methods;

lower: the relative difference from the standard method;

The data for each year are derived from the census and sampling data published by the National Bureau of Statistics printed in the statistical yearbook.

As shown in Figure 1, the total birth rate of China from 2000 to 2016 is calculated by the standard method, the five-age method and the total birth number method, and the difference between the five-age method and the standard method is very small, and the relative difference is mostly within 1% between the five-age method and the standard method, indicating that the use of the five-age method to calculate the total fertility rate is rather accurate. The relative difference between the total birth number method and the standard method is relatively larger, reaching a maximum of 11%, but in most years, the relative error is less than 5%, especially in the census year, when the sample size of the survey is large, both the absolute and relative difference is also much reduced.

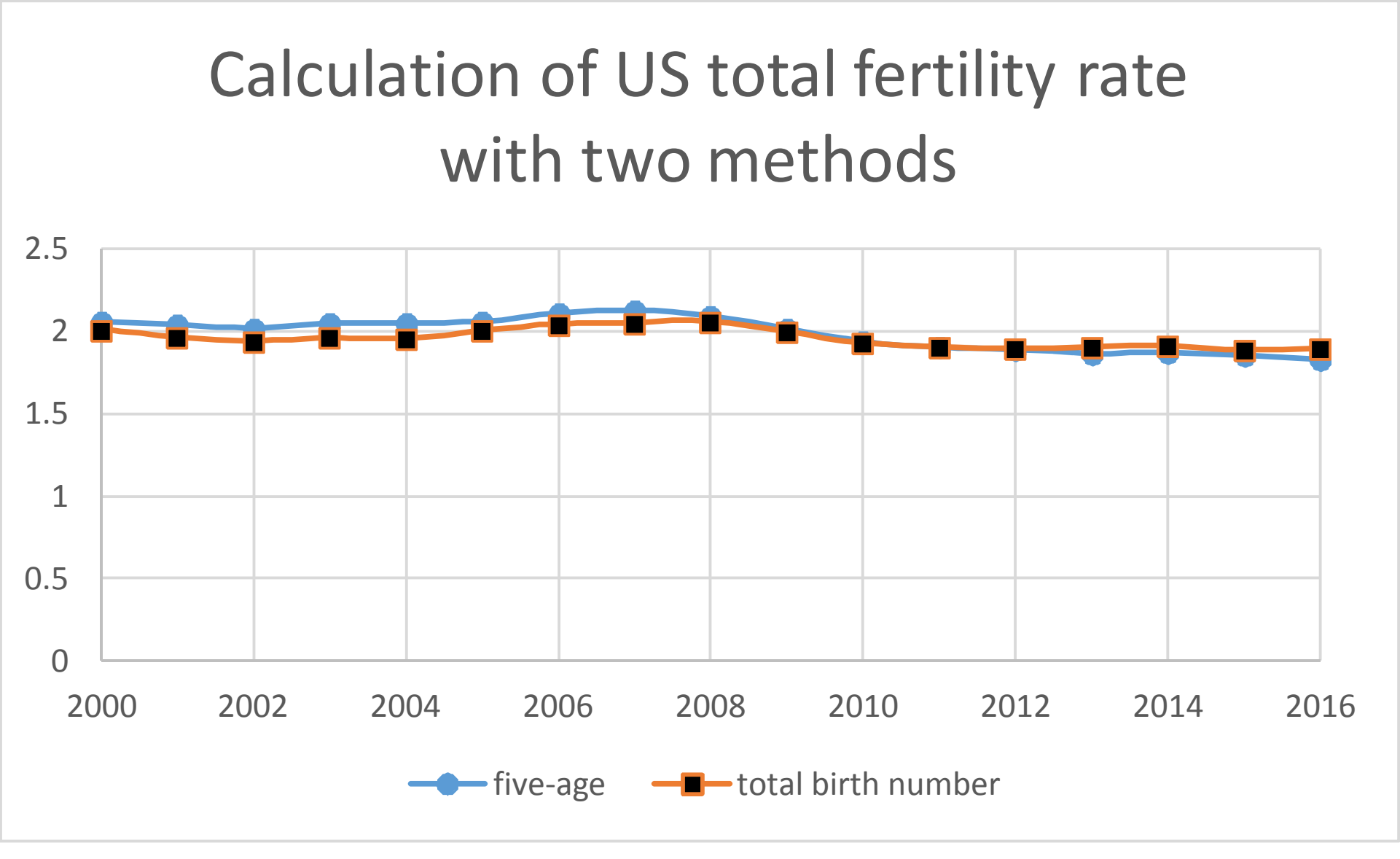

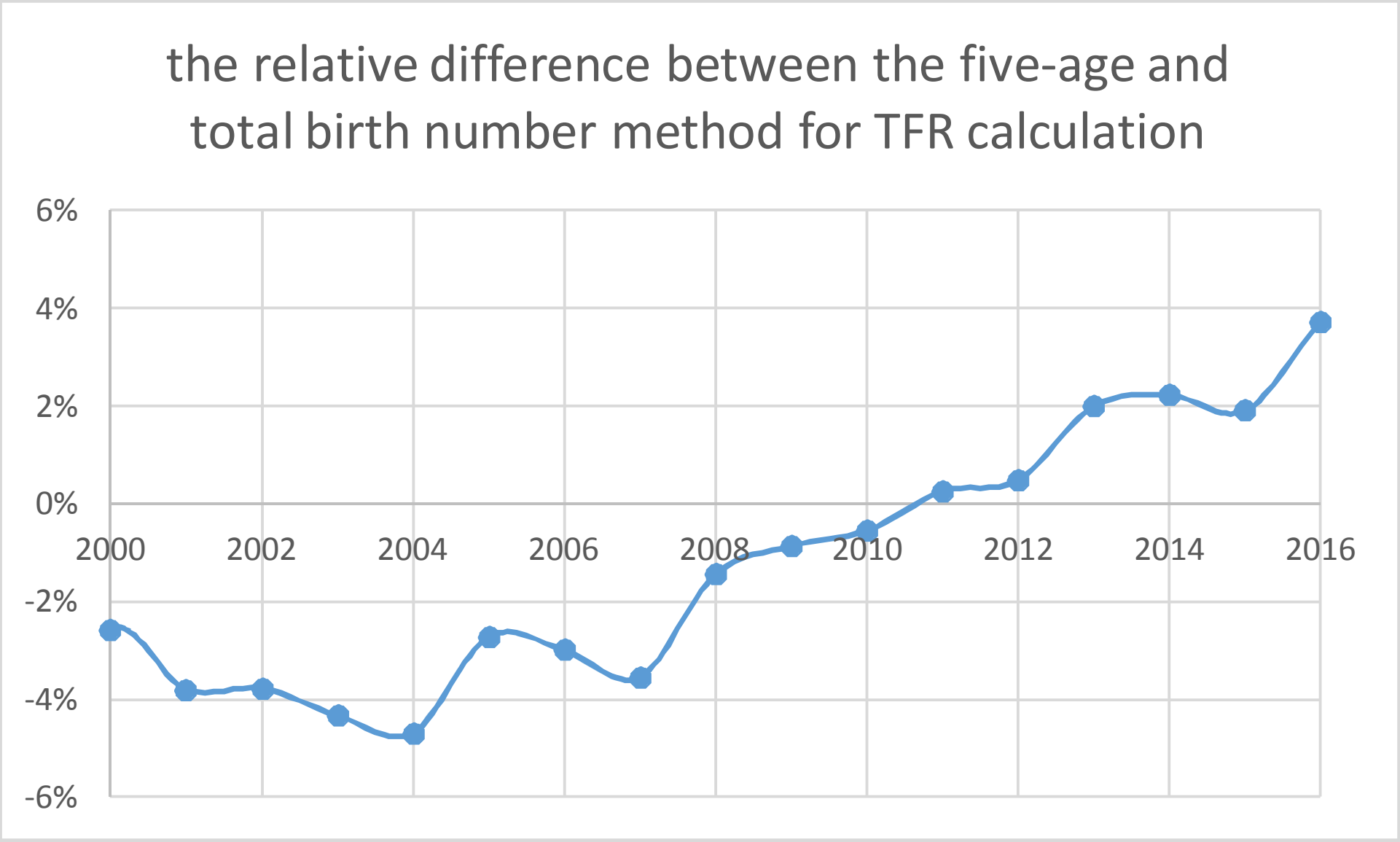

Figure 2. Comparison of the five-age method and the total birth number method to calculate the total fertility rate in the United States.

Upper: Total fertility calculated by the two methods; Lower: Relative difference; Data for each year is from Vital Health Stat reports published by the Centers for Disease Control and Prevention (CDC) of the US Department of Health.

Since there are no statistics on the age-specific fertility rate in all ages in the United States, only the five-age method and the total birth number method can be used and compared. As shown in Figure 2, the absolute difference in the total fertility rate calculated using the two methods is very Small, all less than 0.1, the relative difference is also very small, which is no more than 5%. The results shows that the total birth number method can be used to calculate the total fertility rate with high accuracy.

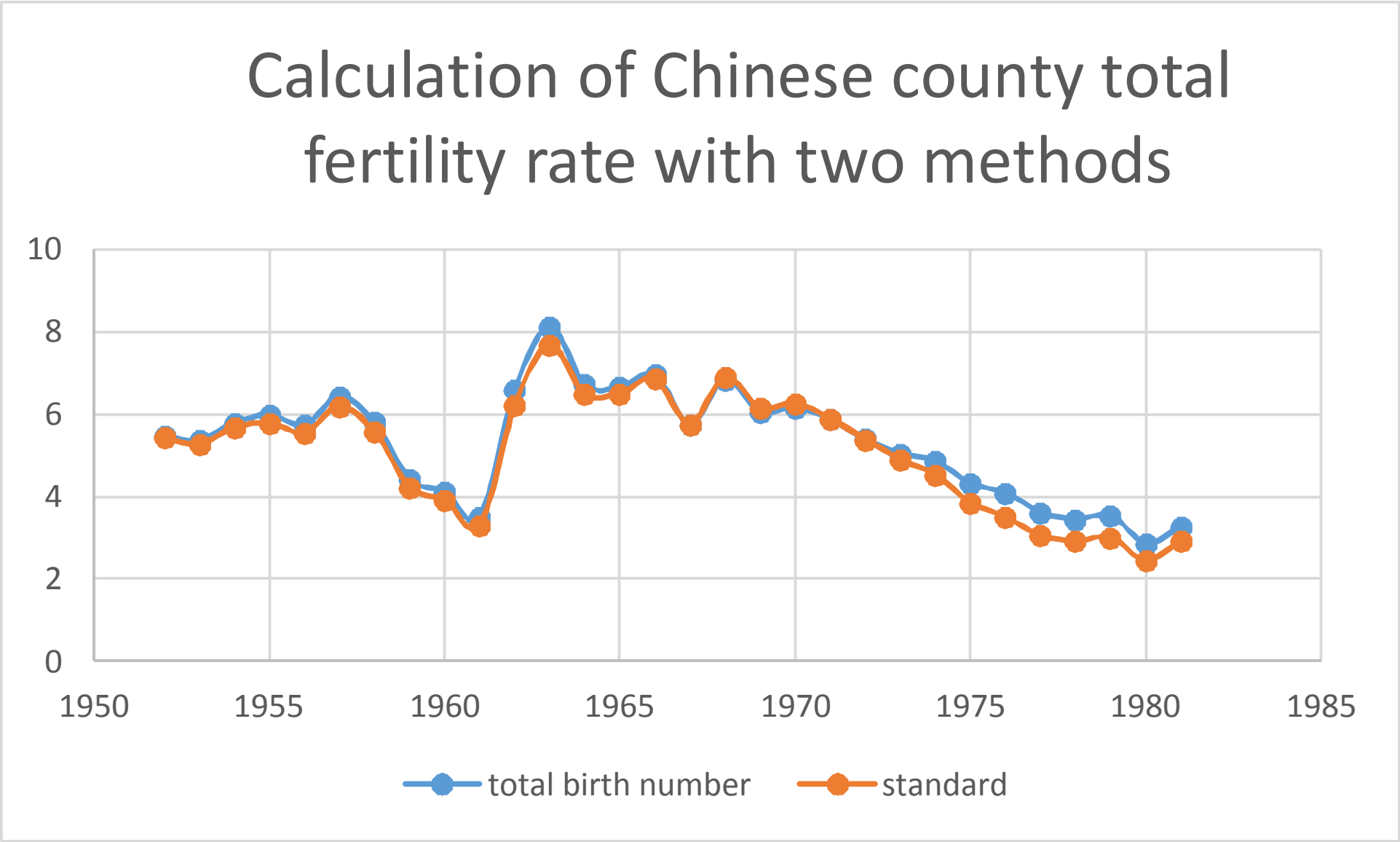

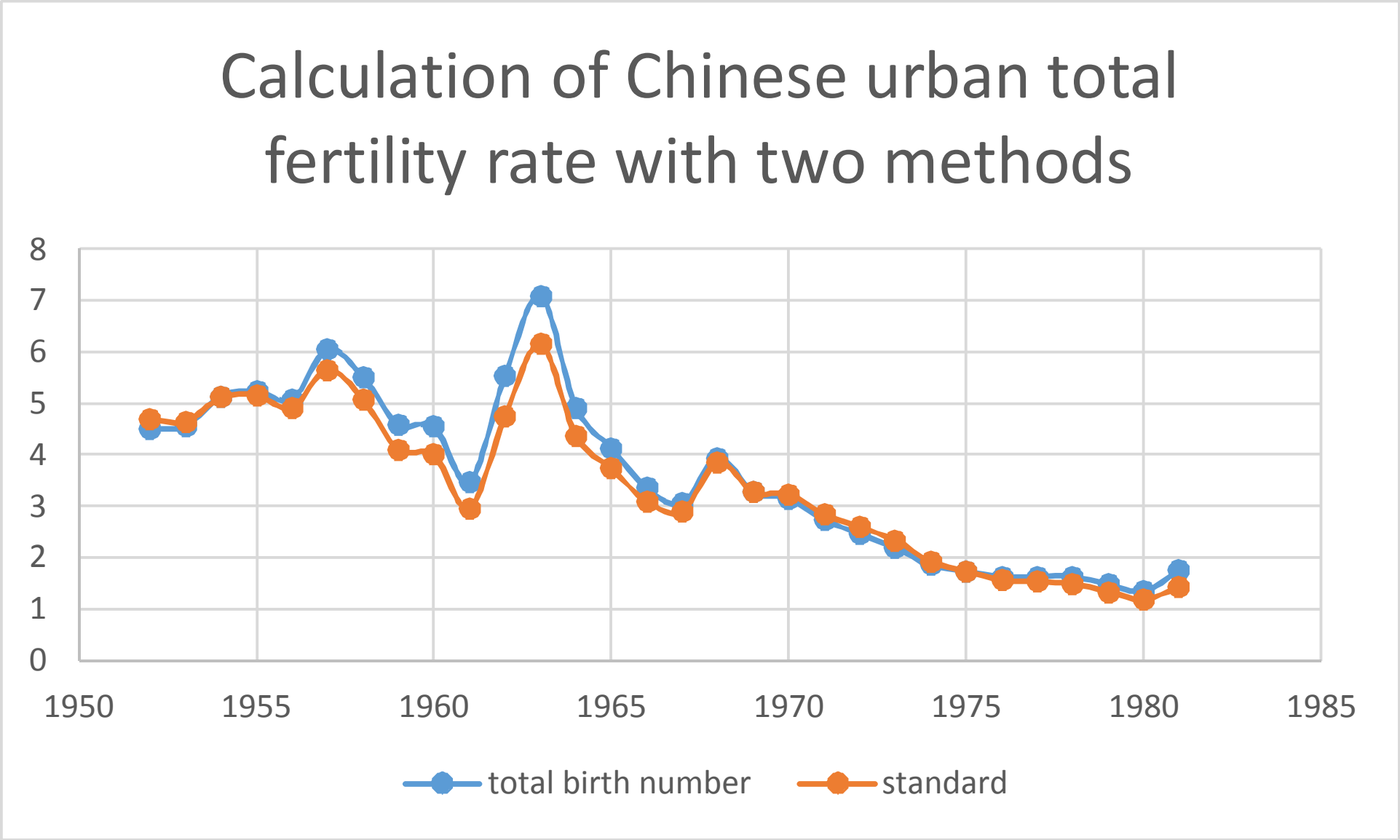

Figure 3. Comparison of standard and total birth number method to calculate the total urban and rural fertility rate in China from 1952 to 1981

Upper: rural; lower: city;

The data of each year comes from the 《Sampling Survey Data of Fertility Rate of One Thousand National Population in China》 published by the National Bureau of Statistics in 1983.

Figure 3 shows the total urban and rural fertility rates in China from 1952 to 1981

using the standard method and the total birth number method. Their absolute differences are also small. These results all indicate that the total birth number method proposed in this paper calculates the total fertility rate, and its error is rather small, which is an acceptable calculation method.

**4 Analysis of Error Sources of Total Birth Number Method**

In principle, both the five-age method and the total birth number method combine with the multiple ages women data. The combined multiple-age women population and the number of birth are used to calculate their multiple-age fertility rate instead of calculating the fertility rate of each age. The simplest combination is to merge two age groups. According to the definition of total fertility, TFR of women in both age groups is

$$TFR=TFR_1+TFR_2==B_1/W_1+B_2/W_2 \quad (7)$$

The combined fertility rate calculated after the merger is:

$$TFR=2*(B_1+B_2)/(W_1+W_2) \quad (8)$$

Assuming that the average population of women of childbearing age in both ages is W, the above equation can be simplified as:

$$TFR=(B1+B2)/W \quad (9)$$

Assuming W2>W1, then d=W-W1=W2-W, substituting (7)

$$TFR=B1/(W-d)+B2/(W+d);$$

Then the difference of the TFR calculated by the two methods dTFR

$$dTFR= [B1/(W-d)+B2/(W+d)] - (B1+B2)/W$$

$$=TFR*(d^2/W^2)/(1- d^2/W^2)+d*(B1-B2)/(W^2-d^2) \quad (10)$$

The relative difference is

$$dTFR/TFR=(d^2/W^2)/(1- d^2/W^2)+d*(B1-B2)/(W^2-d^2)/TFR \quad (11)$$

When the difference of the population number of the two age groups is small, the d/W is very small, and the two terms at equation 11 are small and the relative error will be small. The first term of the two errors is always a positive number, and in some cases, the latter term is negative, and the two will partially cancel each other to reduce the error.

The five-age method and the total birth number method combined the five age populations and the 35 age populations, respectively, and the errors have the similar tendency. This shows that the difference in population size among women of childbearing age will be relatively small when the population is relatively stable for a long time. Therefore, the total fertility rate obtained by the five-age method and the total birth number method will be similar to the standard method, and the relative error will be small, but when the number of women of different ages in childbearing age varies greatly, the relative error of the total birth method will increase.

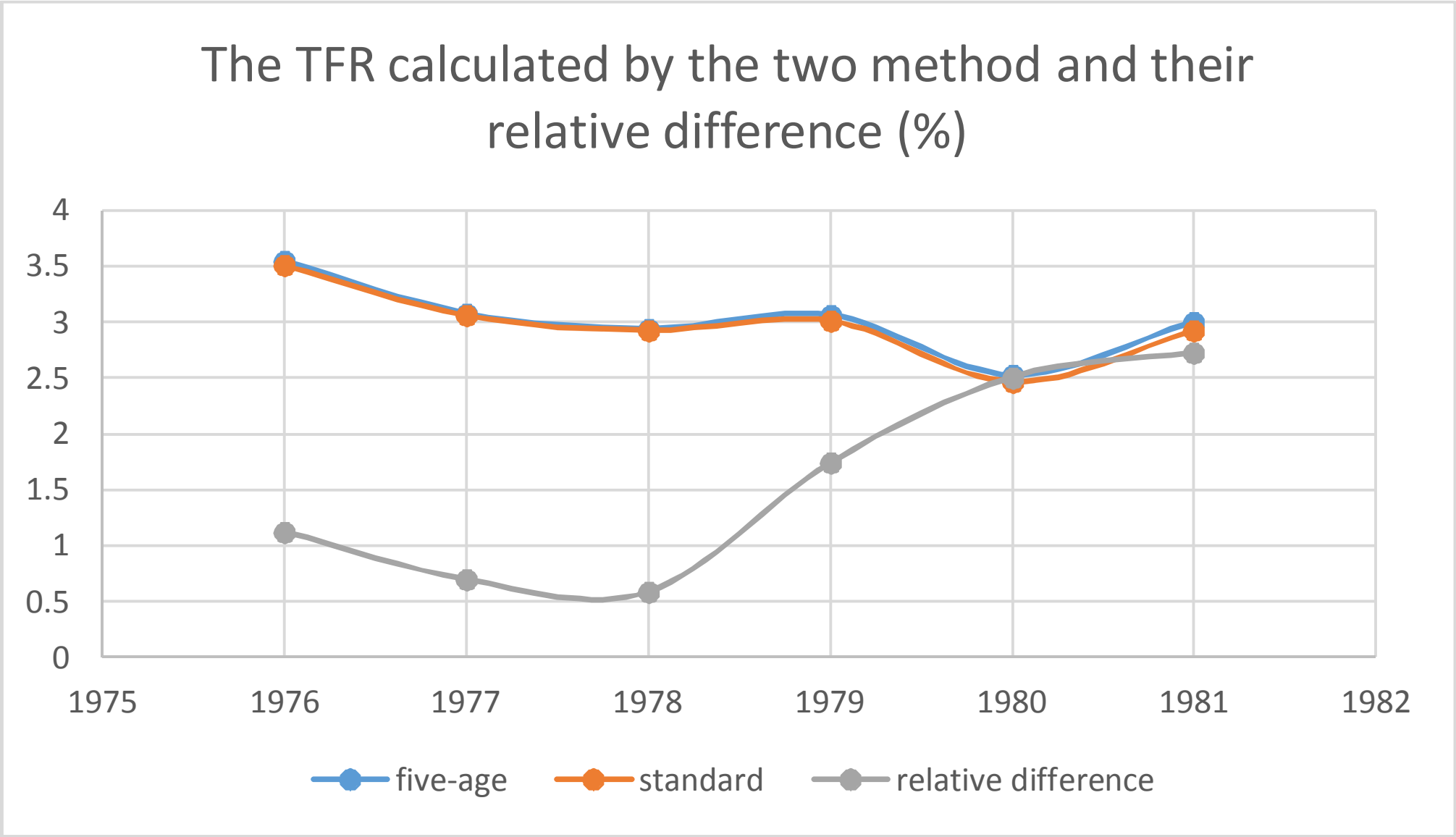

Figure 4, the five-age method and the standard method calculate the total rural fertility rate and their relative difference in China from 1976 to 1981. The data source is the same as figure 3.

As shown in Figure 4, the five-age method and the standard method are used to calculate the total rural fertility rate and relative difference in China from 1976 to 1981. Although the error of the five-age method is small, the relative error exceeds 2.5% for the five-age method which is used to calculate the rural areas in China from 1980 to 1981. The main reason is that in 1980 and 1981, the 20-30 years old women who entered the peak bearing age included women born in 1960 and 1961, their population number is much smaller than those born in the previous years. For example, in the population survey, the number of women born in 1957 was more than 7,900, and in 1963 was 11,885, while the number of women born in 1960 and 1961 was only 4,870 and 4,579. Therefore, even if the five-age method is used to calculate the total fertility rate in 1980 and 1981, the relative error is more than 2.5%. This also shows that the five-age method and the total birth number method are suitable for the calculation of countries and regions with relatively stable population, especially the number of the female population at the childbearing age, which is relatively stable. if the number of the populations at the peak bearing age varies greatly, especially in the 20-30 age group, the error will increase. Using the total birth number method to calculate the total fertility rate in 2006, the relative error is more than 10%. One important reason is that among the women of childbearing age in 2006, the number of 21-year-old population is only 7041, which is much less than the population number of the other age. For example, the female population number of 24-year-old is 8,669, and the number of 17-year-old woman is 10,232.

## 5 Conclusion

This paper deduces a new method for calculating the total fertility rate, which only requires the number of births and the total number of women of childbearing age in the year. In this paper, the age-specific fertility data of China and the United States are used for validation of the total birth number method, and for estimating the error of the method using the standard method and five-age method respectively. It calculates the relative error, which is less than 5% in most cases. However, when the number of women at childbearing age varies greatly between different age groups, the error increases, possibly exceeding 10%.

The calculation equation of the total fertility rate proposed in this paper is a deterministic calculation equation, and it is not necessary to determine the empirical parameter, so it has better application range. This method requires less high-quality data, which provides a simpler method for obtaining a more accurate total fertility rate although some error may be produced. Historically, the number of births was not counted according to the age of the mother(Preston et al. 2001), so it is difficult to estimate the total fertility rate with the standard method or the five-year method, but it can be estimated using the present method. Therefore, the present method offers the possibility of calculating the total fertility rate in the history. It should be pointed out that this method does not solve the underestimation of the total fertility rate caused by the underreporting of the birth population, but the high quality data required is greatly reduced, thus providing an easier way to obtain a more accurate total fertility rate.

From the present method, it also shows that the TFR is proportional to the crude birth rate, the meaning of the proportional coefficient is also given, which proposes a good method to estimate the coefficient, so the method for calculating TFR from the crude birth rate (QIAO and ZHU 2018) is changed from the empirical method to a general method.